\documentclass[sigconf,nonacm]{acmart}

\usepackage{graphicx}
\usepackage{subfiles}
\usepackage{soul,color}

\usepackage[disable]{todonotes}
\usepackage{booktabs}
\usepackage{adjustbox}
\PassOptionsToPackage{hyphens}{url}\usepackage{hyperref}

\usepackage{dsfont}
\usepackage{amsmath}
\usepackage{nicefrac}

\usepackage{amsthm}

\theoremstyle{definition}

\usepackage{caption}
\usepackage{subcaption}
\usepackage{tablefootnote}
\usepackage{pifont}

\usepackage{multicol}
\usepackage{multirow}
\usepackage{url}

\usepackage{tikz}
\tikzset{
  treenode/.style = {shape=rectangle, rounded corners,
                     draw, align=center,
                     top color=white, bottom color=white},
  root/.style     = {treenode, font=\Large, bottom color=red!30},
  env/.style      = {treenode, font=\normalsize},
  dummy/.style    = {circle,draw}
}
\usepackage{diagbox}

\usepackage{xcolor}
\newcommand\eugene[1]{#1}

\AtBeginDocument{%
  \providecommand\BibTeX{{%
    \normalfont B\kern-0.5em{\scshape i\kern-0.25em b}\kern-0.8em\TeX}}}

\begin{document}

\title[Overview of the TREC 2025 RAGTIME Track]{Overview of the TREC 2025 RAGTIME Track 
}
\author{Dawn Lawrie,$^\dagger$ Sean MacAvaney,$^\ddagger$ James Mayfield,$^\dagger$ \\ 
Luca Soldaini,$^*$ Eugene Yang,$^\dagger$ Andrew Yates$^\dagger$}
\affiliation{
  \institution{$^\dagger$Johns Hopkins University Human Language Technology Center of Excellence,\\
  $^\ddagger$University of Glasgow,  $^*$Allen Institute for AI}
  \country{}
}

\email{lawrie@jhu.edu,sean.macavaney@glasgow.ac.uk,mayfield@jhu.edu}
\email{lucas@allenai.org, eugene.yang@jhu.edu, andrewyates@jhu.edu}

\renewcommand{\shortauthors}{Lawrie et al.}

\begin{abstract}

The principal goal of the RAG TREC Instrument for Multilingual Evaluation (RAGTIME) track at TREC
is to study report generation from multilingual source documents.
The track has created a document collection containing Arabic, Chinese, English, and Russian news stories.
RAGTIME includes three task types: Multilingual Report Generation, English Report Generation, and
Multilingual Information Retrieval (MLIR).
A total of 125 runs were submitted by 13 participating teams 
(and as baselines by the track coordinators)
for three tasks.
This overview describes these three tasks and presents the available results. 
\end{abstract}

\settopmatter{printfolios=true}
\maketitle

\section{Introduction}

This is the first year of the RAG TREC Instrument for Multilingual Evaluation (RAGTIME) track at TREC.\footnote{\url{https://trec-ragtime.github.io/}}
RAGTIME provides an opportunity to study Retrieval-Augmented Generation (RAG) systems in the context of a long-form report generation task where systems create reports in one language that summarize the content of documents in several languages.

The RAGTIME report generation tasks are intended to assess how well systems can produce a report detailing relevant facts from retrieved documents. 
These tasks feature detailed report requests consisting of both a problem statement and a user background, a document collection covering four languages, and an emphasis on evaluating whether each claim in a report is supported by a citation into the document collection.
Task participants are provided with a set of detailed report requests in English and asked to generate an English human-readable report that discusses relevant information from retrieved documents and correctly cites those documents in the report. Report requests consist of both a problem statement and a background. Depending on the task version, documents are either in Arabic, Chinese, English, and Russian (Multilingual Report Generation) or only in English (Monolingual Report Generation).

RAGTIME 2025 also offers a Multilingual Information Retrieval (MLIR) task.
In this supporting task, participants are provided with a report request and asked to return a ranked list of documents relevant to the report.
The purpose of this task is to encourage RAG-specific retrieval research and to enrich judgment pools to improve the reusability of the RAGTIME collection. 

\eugene{
RAGTIME 2025 also aims to study the reusability of report generation evaluation.  
We experiment with the reliability of the automatic evaluation process (Auto-ARGUE~\cite{autoargue}) that replicates the ARGUE evaluation framework~\cite{mayfield2024evaluation} that we used to develop our assessment process. 
Different from other TREC tracks, we duplicated topics with two report lengths (2,000 and 10,000 characters) and acquired human annotation without any LLM-in-the-loop to avoid any biases on short reports. 
We use human-curated annotation artifacts to ground the automatic evaluation system for evaluating the long reports. 
These setups allow us to study the reliability, robustness, and generalizability of our assessment process. 
}

The remainder of this paper is organized as follows.
We begin with a summary of the Report Generation and Multilingual Information Retrieval tasks, including a detailed description of the document collections, the 2025 report requests, and the assessment process.
This is followed by an overview of results from the thirteen participating teams.
Finally, we discuss future directions, including what will change in RAGTIME 2026, and conclude.

\section{Multilingual Report Generation}
\label{sec:mlrag-task}
Multilingual information retrieval (MLIR) solves one problem---it ranks documents relative to a query in other languages;
but it creates another---someone has to read all those documents!
This is not just a matter of the time and effort required---some searchers may also not be able to read documents in their original language. 
The goal of the Multilingual Report Generation task is to address both of these challenges
by creating concise focused reports
(i.e., multi-document summaries)
in the language of the report request (which in our case is English).
Each report is based on documents from four languages in the RAGTIME collection (Arabic, Chinese, English, and Russian). 
These reports are evaluated based on the degree to which they use correctly cited references
to documents in the specified collection to answer questions that the report requester wished answered
using the procedure proposed by \citet{mayfield2024evaluation}.
To encourage participation and support the Multilingual Report Generation task, RAGTIME also offers a Monolingual Report Generation task and a Multilingual Information Retrieval task.

\subsection{Monolingual English Report Generation}
RAGTIME provides a monolingual version of the report generation task for participants who prefer to focus on report generation without the need to handle multilingual documents.
This task uses an English subset of the RAGTIME collection.
Otherwise, it is identical to the Multilingual Report Generation task.

\subsection{Multilingual Information Retrieval (MLIR)}
This task expects systems to search all four document collections and produce a single unified ranked list.
This task builds on the Multilingual Information Retrieval tasks at NeuCLIR 2023 and 2024.
However, unlike the MLIR tasks at NeuCLIR, the RAGTIME MLIR topics consist of the entire report request rather than a traditionl TREC title, description, and narrative.
This aligns the topics with the report requests used in the Report Generation tasks.

\section{Documents}

All tasks used the same RAGTIME1 collection,
which contain CommonCrawl News articles\footnote{\url{https://commoncrawl.org/2016/10/news-dataset-available/}} in Arabic, Chinese, English, and Russian. In the case of the Monolingual English task, only the English subset of the collection is used.

The documents were obtained by the CommonCrawl service between August 1, 2021 and July 31, 2024.
Text was extracted from each source web page %
Language id performed by GlotLID v3, which the FineWeb folks at HuggingFace recommended.
Short and long documents were filtered from the collection.
Short documents were determined by token count. Documents under 100 tokens based on the GPT-4 tokenizer were removed.
Long documents were determined by character count.
The maximum number of characters in a document is 24,000 characters. 
The crawl date was associated with each document, which given the timeframe of the collection is generally very close to when the document was published.
In order to balance the collection, a sample of 915 documents per day per language was taken. This lead to just over one million documents in the collection per language. 
Final collection statistics appear in Table~\ref{tab:coll_stats}.

RAGTIME1 can be downloaded from Huggingface Datasets.\footnote{\url{https://huggingface.co/datasets/trec-ragtime/ragtime1}}
 The collection is distributed in JSONL,
a list of JSON objects, one per line.
Each line represents a document.
Each document JSON structure consists of the following fields:
\begin{description}
\itemsep0em 
    \item [id:] a unique string for the document %
    \item[time:] crawl time 
    \item[text:] article text
    \item[url:] source url for the document when it was crawled
\end{description}

\begin{table*}[tb]
\centering
\caption{Document Collection Statistics for RAGTIME1 (token counts GPT-4 tokenizer)}.\label{tab:coll_stats}
\begin{tabular}{c|c|cc|cc}
\toprule
         & Document & Avg. Chars   & Median Chars & Avg. Tokens  & Median Tokens \\
Language & Count    & per Document & per Document & per Document & per Document  \\
\midrule
Arabic  & 1,000,095 & 1740 & 1355 & 545.43 & 424 \\
Chinese & 1,000,095 &  849 &  694 & 730.98 & 606 \\
English & 1,000,095 & 3100 & 2478 & 683.03 & 544 \\
Russian & 1,000,095 & 1660 & 1191 & 489.33 & 343 \\ 

\bottomrule
\end{tabular}
\end{table*}

\section{Report Requests}

A report request consists of a request ID, a collection ID, a title, a background section, a problem statement,
and a length limit (in Unicode characters).
The title, background, and problem statement fields of a report request are expressed in unstructured text.
Here is an example:
{ \ttfamily

\noindent Title: Machu Picchu Architecture

\noindent Background: As an archaeologist leading an expedition in South America, I require insights into the "Mysteries of Machu Picchu's Architecture" to deepen our team's understanding of its construction techniques and histor- ical significance. This report will guide our fieldwork and contribute to the scholarly discourse on Incan civilization.

\noindent Problem statement: Produce a report on the mysteries of Machu Picchu's architecture. The focus of the report is on speculations and theories regarding the construction methods and architectural marvels of Machu Picchu. I am also interested in hypotheses about the purpose, techniques, and significance of the unique structures at this ancient Incan site.

\noindent Limit: 2000
}

NIST Assessors worked in pairs to create report requests. 
The goal was to create a request where relevant nuggets could be found in documents from at least two of the four languages.
Unlike developing topics for retrieval tasks, we decided that it would not be a problem if the report request had lots of relevant documents because we were not concerned about judging all relevant documents. Whenever a person writes a report, judgments must be made about what information is important enough to include in the report and what gets left out. Assessors are expected to make such decisions based on the background that describes the person with the request.  
After drafting a problem statement, each assessor was asked to find at least five documents containing nuggets and search for nuggets in documents written in two of the four languages.
Each assessor could read one of the non-English
languages. For that language, they read the documents in the language in which they were written. Their second language was either English (all assessors could also read English) or machine translations of the fourth language.

In each language, at most 30 documents were examined. Documents could be retrieved based on the title or assessors could search for particular facets of the problem statement using multiple queries. Once the assessor found a document containing a vital or okay nugget, they were instructed to move to their other language and read no more than 30 documents. For each document they assessed whether or not it contained a vital or okay nugget. To complete the annotation, two of four languages were required to have documents containing vital nuggets and there to be a total of 10 documents containing vital or okay nuggets.

Assessors created report requests in two batches. Twenty-five requests were distributed as the dry-run requests. There were twenty unique problem statements. Five of the requests had the same problem statement but different backgrounds. All requests had a limit of 2,000 characters.
The main task consisted of fifty unique problem statements. Eleven requests had the same problem statement with two different backgrounds, leading to sixty-one requests. Each request was paired with a short limit of 2,000 characters and a long limit of 10,000 characters. Therefore, the request file consisted of 122 requests.
\section{The Assessment of Runs}

\begin{table*}[t]
\caption{Initial Subset of Raw Report Generation Assessment Statistics. We report the statistics on the short topics only in the main task.}\label{tab:repgen-assessment-stats}
\centering
\begin{tabular}{l|cc}
\toprule
{}                             & Dry-Run & Main  \\
\midrule
\# of Report Request Developed & 25 & 61 \\
\midrule
\# of Report Request Assessed & 18 & 34 \\
\# of Report Request Assessed with Multiple Backgrounds & 3 & 4 \\
Avg. \# of Nugget Questions per Request  & 146.1  & 16.0  \\
\bottomrule
\end{tabular}

\end{table*}

NIST assessors created the ground truth data that was used to evaluate Report Generation runs.
Our goal was to create data that would support the evaluation design described by \citet{mayfield2024evaluation}. For both the dryrun and main task, assessors reviewed documents and marked passages that contained useful information in Phase~1.
For the main task, assessors continued to Phase 2 where they defined high-level questions, and the low-level Question/Answer pairs needed for ARGUE assessment.
Phases 3 and 4 produce robust data for future use of the evaluation data and manual evaluation of the generated reports submitted by participating teams. In Phase 3, assessors review sentences that cite a document to determine whether the information attested in the sentence appears is attested in the document. 
In Phase 4, assessors determine which nugget questions are addressed by a generated report.
A summary of the assessment statistics is presented in Table~\ref{tab:repgen-assessment-stats}. %

\subsection{Phase 1: Document Relevance}

The first step to judge document relevance was to create document pools to judge. Since relevance labels are applied to problem statements, rather than report requests with their background and length limit, submissions for all task across any request with the same problem statement were combined to create one document pool per problem statement.

The documents in the pool combined the top three cited documents for each report sentence in a generated report as well as the top three documents in the retrieval runs. Then documents were divided by language. Assessors in each language reviewed all documents. One of the assessors for that topic was also assigned the English set of documents. They applied a 4-level scale when judging documents:
\begin{itemize}
    \item 
\textbf{Very valuable} [3 pts]: documents that contain information that is central to the topic and is
something that would be put into the lead paragraph of a report. It is an excellent citation for a report.
\item
\textbf{Valuable} [1 pt]: documents that include information that is central to the topic and 
 the information in the document would be included in the report, but not as prominently.
\item
\textbf{Topical} [0 pts]: documents that just mention or touch on the topic, even a single sentence.
\item
\textbf{Irrelevant} [0 pts]: documents that do not relate to the topic at all.
Documents that only have relevant information in "teaser" links at the bottom of the
article ("More articles you might be interested in" kinds of links) are not relevant. \end{itemize}

The points appearing with the label are the values assigned in the qrels for evaluating retrieval tasks. Notice that while assessors have 4-levels, the qrels will only have three different point values since topical and irrelevant are both mapped to 0 points.

In order to support report generation assessment, assessors also highlighted passages in the document the first time they encountered the information.
The goal of this task was to assemble a list of relevant facts from the documents to use for
writing a report without having to read the document a second time. While these passages were used as the source of information to create a set of automatically generated nuggets in the dryruns, in the main task assessors translated the salient facts into English in such a way that the information could be used on its own by any assessor to compose nugget question-answer pairs about the information that should appear in a generated report. This enabled a single opinion about the required content of the generated report without requiring one person to have language skills in all of the languages represented in the document collection.

In the dryruns, assessment stopped after this phase. All other information necessary for scoring a generated report was created with LLMs. Thus the highlighted passages lead to over 140 nugget questions on average as is shown in Table~\ref{tab:repgen-assessment-stats}.

\subsection{Phase 2: Nugget Creation}
The purpose of this phase is to identify the high-level questions that a good report on this problem statement given the background of the requester would want to answer. Then facts that address this high-level question are selected from the list of facts identified during Phase 1. The important pieces of information expressed in those facts are rewritten as one or more nugget question and answer pairs that will be used to assess the nugget coverage of the generated report. 

\subsection{Phase 3: Citation Assessment}

\eugene{
The purpose of this phase is to determine whether each generated report sentence is
supported by the document that is cited as providing support for the sentence. 
Each report sentence and cited document pair is grouped by the document and split by the language in order to be assigned to the assessors with the corresponding language expertise. 
We primarily assign pairs of a topic here to the same assessors in Phase 1, but not always because of assessors' availability and their annotation pace. 
}

\eugene{
Each report sentence and document pair is judged for one of ``fully supported'', ``partially supported'', and ``not supported''. 
For evaluation, we only consider a sentence to be supported if it is \textit{fully} supported by each of its cited documents \textit{individually}, i.e., a sentence is considered not supported if it is not fully supported. A document that partially supports a sentence is not included because it is unknown what information in the sentence is not supported by the document.
Across all languages and runs, 25,032 annotations are needed. However, only 17,071 were acquired because of an error during task dispatch. The remaining 7,961 were filled with an LLM-as-a-judge using OpenAI GPT-5 Mini. 
We use the same model and prompt to judge a balanced sample of 1,000 judged pairs for validation, and the model reaches a precision and recall of 0.815 and 0.814. 
While this accuracy is not as perfect as we would like, it is similar to the agreement when the coordinators manually checked a small sample of assessments. 
More investigation into the annotation process and human validation is required to assess how reliable the labels are. 
We discuss the impact of the LLM-filled labels in scoring later in the results section. 
}

\subsection{Phase 4: Nugget Matching}

\eugene{
The purpose of this phase is to determine which nuggets are included in each generated report.
Assessors assess each report sentence against each nugget created in Phase 2. 
A total of 24,399 report sentences have been assessed, and 9,485 of them \textit{mentioned} at least one nugget. 
Since all nuggets are already in English, this phase does not require specific language expertise like those in Phases 1 and 3. 
To avoid the assessors learning new topics, we still assign Phase 4 annotation to assessors who have worked on the topic already in at least one of the previous phases. 
}

\eugene{
However, a nugget is considered \textit{covered} only if the sentence is also grounded in its citations, which inherit the LLM-filling noise from sentence support assessment in Phase 3. 
We will also discuss the impact of such noise in the next section. 
}

\subsection{Evaluation}

Topics with a 2000-character limit (short topics) received a *manual* evaluation based on the ARGUE framework~\cite{mayfield2024evaluation}, with manual nugget alignment and sentence-citation using manual support assessment \textit{when available}. In the rare cases where a citation was unjudged, an automated prompt-based classifier tuned with a sample of judged data filled in the missing judgment leading to mostly manual evalation.   
Topics with a 10000-character limit (long topics) are evaluated with Auto-ARGUE~\cite{autoargue}, an automatic implementation of the ARGUE framework, using the corresponding human-curated nuggets. 
We use the curated nuggets as the nuggets for automatic evaluation for all short topics, as well as a validation of the automatic evaluation, since we do have mostly manual evaluation on them.
All nuggets are tagged as OR nuggets, i.e., covering any answer would receive full credit for the nugget. 
We use a Llama3 70B Instruct model as the backbone for the automatic evaluation. 

\eugene{
We primarily report scores on the following metrics
\begin{itemize}
    \item sentence support: a precision-based metric that reports the proportion of the report sentence that are supported by its citation. Note that a sentence is considered supported only when all of its citations support the sentence independently. 
    \item nugget coverage: a recall-based metric that reports the proportion of the nuggets that are correctly mentioned and grounded in citations. 
    \item F1: the harmonic mean of the sentence support and nugget coverage. 
\end{itemize}
For each run, report the macro-average values across topics. 
While reporting the same metrics, since short and long topics are generally based on different evaluation processes (manual vs automatic), we report short and long topic macro-average values separately. 
}

\section{Additional Resources}

\subsection{Retrieval Service}

To support teams primarily focusing on the report generation task,
we provided a PLAID-X~\cite{tdistill} search service through a web API
that used an English-trained model~\cite{yang2024distillation}\footnote{\url{https://huggingface.co/hltcoe/plaidx-large-eng-tdist-mt5xxl-engeng}} for all languages. 
To minimize the resources needed to host the service,
we included the ability to remove documents in other than the requested language. 
The user can request up to 100 documents for each query.
The service retrieves ten times the number of documents the user requested
to ensure enough documents in the requested language are retrieved. 

\subsection{Development Data}

The NeuCLIR 2024 Report Generation Pilot~\cite{lawrie2024overview} was available as development data.
This data consists of fifty-nine report requests, twenty-two of which have QCed nugget question-answer pairs with answers linked to documents in at least one of the three languages in the NeuCLIR1 collection.
Nineteen of those report requests have nugget answers linked to documents in all three languages: Chinese, Persian, and Russian.
These nineteen topics are suitable development data for the Multilingual Report Generation task.

\section{Participation}

\begin{table*}[t]
    \centering
    \caption{Summary of participating teams. }
    \label{tab:participants}
\begin{tabular}{ll|cc}
\toprule
          &         & Report Gen  & MLIR \\
Team Name & Team ID & Submissions & Submissions \\
\midrule
Democritus University of Thrace & DUTH\_XANTHI & 10 & 10 \\
Human Language Technology Center of Excellence & HLTCOE & 10 & -- \\
JHU HLTCOE SCALE25 gen multiagent & hltcoe-multiagt & 10 & -- \\
JHU HLTCOE SCALE25 rerank & hltcoe-rerank & 9 & 10 \\
IDA Center for Computing Sciences & IDACCS & 5 & -- \\
NC State University Laboratory for Analytic Sciences & ncsu-las & 5 & -- \\
Centrl South University & CSU & 3 & -- \\
Adam Mickiewicz University & AMU & 2 & -- \\
DFKI & DFKI & 2 & -- \\
GenAIus Technologies & GenAIus & 2 & 4 \\
University of Amsterdam & UvA & 1 & -- \\
WueRAG & WueRAG & 1 & -- \\
\textit{RAGTIME Coordinators} & coordinators & 1 & 8 \\
\bottomrule
\end{tabular}

\end{table*}

Table~\ref{tab:participants} shows the thirteen teams that participated in the tracks and the number of runs submitted for each task. Teams either participated in the multilingual or English report generation, but not both report generation tasks.
\section{Evaluation Results}

\subsection{Multilingual and English Generation Report Generation}

\subsubsection{Run Statistics}
\begin{table*}[t]
    \centering

    \caption{Report Generation Runs Statistics. The sum of the average citation by language should match the average number of unique citations. If it does not, it indicates some citations are invalid (i.e., document ID not found in the collection). }
    \label{tab:repgen-run-stats}

\begin{tabular}{ll|rrrr|rrrr}
\toprule

Topic &  \multirow[c]{2}{*}{Team ID}  &  Avg. \# &  Avg. \# &  Avg. \#  &  Avg. \# Uniq. & \multicolumn{4}{c}{Avg. \# Citation by Lang}\\
Type  &     &  Char.   &  Sent.   &  Citation &  Citation      & Chinese & Russian & Arabic & English \\
\midrule
\multicolumn{10}{l}{Multilingual Subtask} \\
\midrule
\multirow[c]{9}{*}{Long} 
& AMU & 2796.16 & 15.48 & 22.84 & 5.45 & 0.80 & 0.66 & 0.79 & 3.16 \\
& CSU & 2683.23 & 16.99 & 45.02 & 11.65 & 1.74 & 2.25 & 3.52 & 4.15 \\
& GenAIus & 6311.47 & 30.80 & 30.80 & 13.31 & 3.40 & 1.46 & 2.74 & 5.71 \\
& HLTCOE & 8106.33 & 50.54 & 50.53 & 13.30 & 3.50 & 1.67 & 2.59 & 5.55 \\
& IDACCS & 11662.51 & 85.66 & 85.66 & 27.50 & 6.70 & 7.02 & 6.81 & 6.97 \\
& coordinators & 1319.23 & 18.95 & 32.66 & 10.07 & 2.10 & 1.20 & 1.95 & 4.82 \\
& hltcoe-multiagt & 4936.18 & 32.40 & 34.49 & 7.99 & 1.43 & 0.73 & 1.65 & 4.18 \\
& hltcoe-rerank & 6244.12 & 44.96 & 47.76 & 18.98 & 4.16 & 2.34 & 4.34 & 8.13 \\
& ncsu-las & 5788.73 & 34.47 & 34.47 & 10.23 & 2.63 & 1.17 & 1.87 & 4.56 \\
\midrule
\multirow[c]{9}{*}{Short} 
& AMU & 1644.55 & 10.64 & 15.07 & 5.20 & 0.76 & 0.59 & 0.75 & .07 \\
& CSU & 1160.30 & 7.34 & 24.92 & 11.04 & 1.70 & 2.02 & 3.53 & 3.78 \\
& GenAIus & 1735.05 & 9.05 & 9.05 & 5.47 & 1.46 & 0.70 & 1.02 & 2.29 \\
& HLTCOE & 1856.33 & 12.42 & 12.41 & 6.75 & 1.73 & 0.78 & 1.22 & 3.03 \\
& IDACCS & 2343.70 & 17.61 & 17.61 & 6.38 & 0.00 & 0.00 & 0.00 & 6.38 \\
& coordinators & 1319.23 & 18.95 & 32.66 & 10.07 & 2.10 & 1.20 & 1.95 & 4.82 \\
& hltcoe-multiagt & 1693.62 & 12.04 & 12.84 & 6.01 & 1.08 & 0.55 & 1.19 & 3.19 \\
& hltcoe-rerank & 1977.05 & 16.87 & 19.58 & 12.05 & 2.67 & 1.57 & 2.63 & 5.18 \\
& ncsu-las & 2946.90 & 17.27 & 17.27 & 7.46 & 1.87 & 0.82 & 1.32 & 3.44 \\
\midrule
\multicolumn{10}{l}{English Subtask} \\
\midrule
\multirow[c]{4}{*}{Long} 
& DFKI & 10715.20 & 1.00 & 50.00 & 38.45 & -- & -- & -- & 38.45 \\
& DUTH\_XANTHI & 1065.43 & 5.85 & 4.85 & 3.00 & -- & -- & -- & 1.85 \\
& UvA & 8181.57 & 47.46 & 63.74 & 27.02 & -- & -- & -- & 27.02 \\
& WueRAG & 1173.08 & 7.64 & 9.03 & 4.36 & -- & -- & -- & 4.36 \\
\midrule
\multirow[c]{4}{*}{Short} 
& DFKI & 10761.07 & 1.00 & 50.00 & 38.24 & -- & -- & -- & 38.24 \\
& DUTH\_XANTHI & 1042.86 & 5.82 & 4.82 & 3.00 & -- & -- & -- & 1.85 \\
& UvA & 2065.43 & 11.87 & 15.21 & 8.82 & -- & -- & -- & 8.82 \\
& WueRAG & 1201.28 & 7.70 & 9.18 & 4.34 & -- & -- & -- & 4.34 \\
\bottomrule
\end{tabular}

\end{table*}

Among the 61 report generation runs submitted to RAGTIME, the multilingual report generation task received 47, while 14 are English-only runs. 
Table~\ref{tab:repgen-run-stats} summarizes the statistics of the runs. Most runs exceed the character limit, which is either 2000 or 10000 characters. We truncate the report to the specified limit before evaluation.

\begin{figure*}
    \centering
    \includegraphics[width=\linewidth]{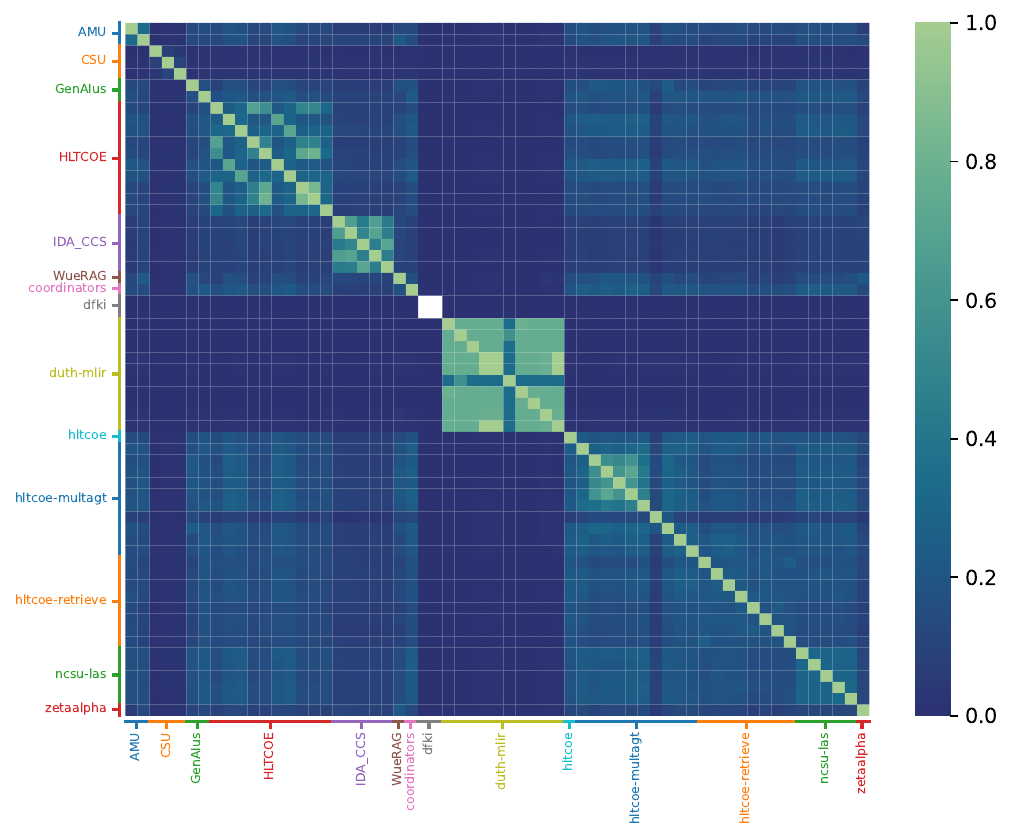}
    \caption{Average citation overlap over all main task (1001 to 1122) topics between all run pairs. White cell indicates no valid topic ID was found in both runs of the pair.}
    \label{fig:repgen-citation-overlap}
\end{figure*}

Figure~\ref{fig:repgen-citation-overlap} shows the citation similarity between all pairs of report generation runs. 
The similarity is calculated by the intersection of the citations of the two runs divided by their union. 
Each team exhibits clear clusters among their submissions, indicating similar retrieval pipelines across the runs. 
Notably, Team \texttt{duth-mlir} runs cited very similar sets of documents, resulting in a clear, bright cluster. 
Other teams, while also similar among themselves, are not as extreme. 

\subsubsection{Human Annotation Results}

\begin{table*}[t]
    \caption{Correlation between LLM-Filled version of the metrics and optimistic and pessimistic versions.}
    \label{tab:human-backfill-corr}
    \centering
    \begin{tabular}{l|ccc|ccc}
\toprule
 & \multicolumn{3}{c|}{Topic-level Pearson Correlation} & \multicolumn{3}{c}{Run-level Kendall's $\tau$} \\
            & Sentence Support & Nugget Coverage & F1  & Sentence Support & Nugget Coverage & F1 \\
\midrule
LLM-Filled  & 1.0000 & 1.0000 & 1.0000 & 1.0000 & 1.0000 & 1.0000 \\
Optimistic  & 0.2089 & 0.8180 & 0.7439 & 0.4824 & 0.8183 & 0.7696 \\
Pessimistic & 0.9822 & 1.0000 & 0.9999 & 0.9053 & 0.9976 & 0.9988 \\

\bottomrule
\end{tabular}

\end{table*}

\eugene{
Tables~\ref{tab:repgen-multi-results} and \ref{tab:repgen-eng-results} summarize the human evaluation and automatic evaluation results on the short topics of the report generation runs, split into the multilingual and English subtasks. 
Here, nugget coverage is particularly challenging for the submitted systems, where the highest nugget coverage value based on human annotations is only 0.227. 
Such low nugget coverage scores result in low overall F1 scores.
Compared to Auto-ARGUE values, assessors are also stingier on awarding sentences being supported by their citations, resulting in lower sentence support scores. 
Interestingly, no system dominates both nugget coverage and sentence support, indicating that there is a large room for improvement in both directions, as well as system design to improve the two aspects simultaneously. 
}

\begin{figure*}
    \centering
    \includegraphics[width=\linewidth]{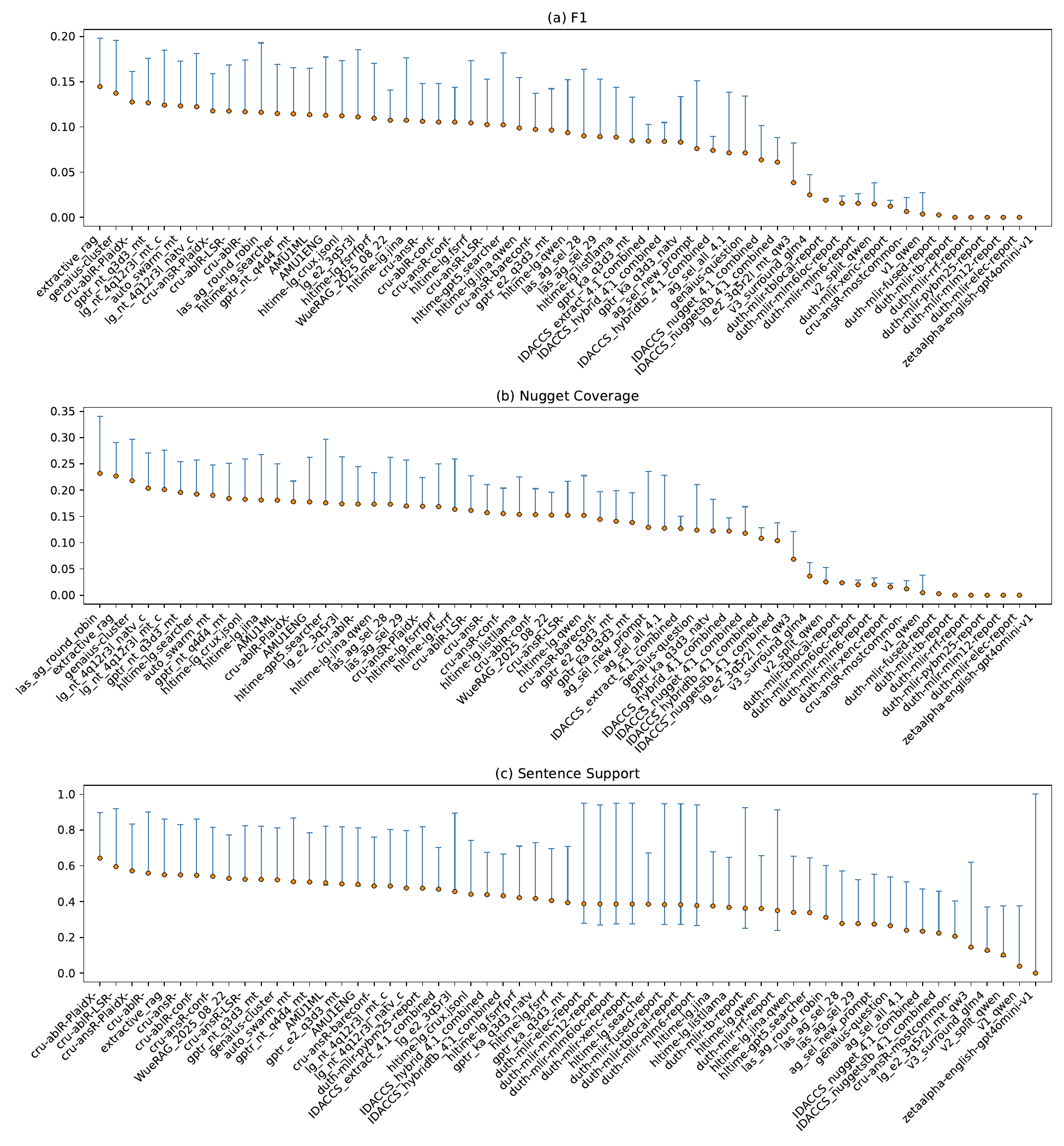}
    \caption{Human optimistic, pessimistic, and LLM-filled metric values of each report generation run on short topics. The top and the bottom whiskers are the optimistic and pessimistic values, with the circles being the LLM-filled values.}
    \label{fig:opt-pes-llmfilled-run-values}
\end{figure*}

\eugene{
Since there are missing assessments in the citation support, we used an LLM-as-a-judge to fill the unjudged instances (LLM-filled version), which introduces LLM noise into the human evaluation results.
To characterize the impact, we create an \textit{optimistic} and a \textit{pessimistic} version of the sentence support metric, which assume the unjudged instances being supported and not supported, respectively. 
We also use the two versions to create an optimistic and pessimistic version of F1 and nugget coverage, since a nugget being \textit{covered} by a report sentence requires the sentence being grounded in the citation as well. 
Table~\ref{tab:human-backfill-corr} summarizes the correlation between the LLM-filled version and the two assumed versions. We can see that the LLM-filled version is nearly perfectly correlated with the pessimistic version, indicating that the LLM judge is also being very strict in awarding a sentence being grounded. 
Figure~\ref{fig:opt-pes-llmfilled-run-values} further illustrates the impact of the LLM judge in each metric and the upper (optimistic) and lower (pessimistic) bounds of the values. 
\todo{I honestly don't have any good intuition on this, but it is fun to stare at these plots. }
}

\eugene{
While it is unfortunate that RAGTIME 25 is not completely human-assessed, the scores are still highly informative for system development. 
}

\subsubsection{Automatic Evaluation with Auto-ARGUE}

\eugene{
Exploring automatic evaluation is a central focus of RAGTIME 2025. 
We evaluated both short and long topics with Auto-ARGUE~\cite{autoargue} using nuggets created by the assessors based on the document pools. 
Table~\ref{tab:repgen-multi-results}, \ref{tab:repgen-eng-results}, and \ref{tab:repgen-results-long} summarize scores produced by Auto-ARGUE on both short and long topics. 
}

\eugene{
To assess the usability and reliability of the automatic evaluation process, we assess the correlation between the human and Auto-ARGUE evaluation scores on the short topics. 
As shown in Table~\ref{tab:human-autoargue-short-corr}, the two evaluation processes correlate strongly on the run level (macro-averaged scores over the topics) with 0.622 Kendall's correlation coefficient on F1, and 0.663 and 0.549 on sentence support and nugget coverage, respectively. 
Given the wide range of scores among the submissions, we also measure the gap-aware rank correlation using $\tau_{GAP}$~\cite{taugap} to discount the penalties of rank swaps when the gaps are small. 
The correlation is even weaker between Auto-ARGUE and human annotation, measured with $\tau_{GAP}$. 
}

\begin{table*}[t]
    \caption{Correlation between Auto-ARGUE and human evaluation on short topics. The diagonal cells of the left and right sections are the correlation between the human and Auto-ARGUE on the same metric. }
    \label{tab:human-autoargue-short-corr}
    \centering
\begin{tabular}{l|ccc|ccc|ccc}
\toprule
& \multicolumn{3}{c|}{Topic-level Pearson Correlation} & \multicolumn{3}{c|}{Run-level Kendall's $\tau$} & \multicolumn{3}{c}{Run-level $\tau_{GAP}$~\cite{taugap}} \\
\midrule
\multirow{2}{*}{\diagbox[height=2em]{Human}{Auto-ARGUE}}
& Sentence & Nugget   &    & Sentence & Nugget   &    & Sentence & Nugget   &    \\
& Support  & Coverage & F1 & Support  & Coverage & F1 & Support  & Coverage & F1 \\
\midrule
Sentence Support & 0.440 & 0.028 & 0.171 & 0.663 & $-$0.035 & 0.216 & 0.521 & $-$0.088 & 0.061 \\
Nugget Coverage  & 0.262 & 0.604 & 0.607 & 0.436 & \phantom{$-$}0.549 & 0.704 & 0.239 & \phantom{$-$}0.417 & 0.549 \\
ARGUE F1         & 0.318 & 0.567 & 0.607 & 0.575 & \phantom{$-$}0.419 & 0.622 & 0.407 & \phantom{$-$}0.322 & 0.496 \\

\bottomrule
\end{tabular}
\end{table*}

\begin{table}[t]
    \caption{Metric correlation between long and short topics using Auto-ARGUE.}
    \label{tab:short-long-corr}
    \centering
\begin{tabular}{l|cc}
\toprule
 & Topic-level  & Run-level   \\
 & Pearson      & Kendall's $\tau$ \\
\midrule
F1 & 0.891 & 0.600 \\
Nugget Coverage & 0.883 & 0.583 \\
Sentence Support & 0.854 & 0.849 \\
\bottomrule
\end{tabular}

\end{table}

\begin{figure*}
    \centering
    \includegraphics[width=\linewidth]{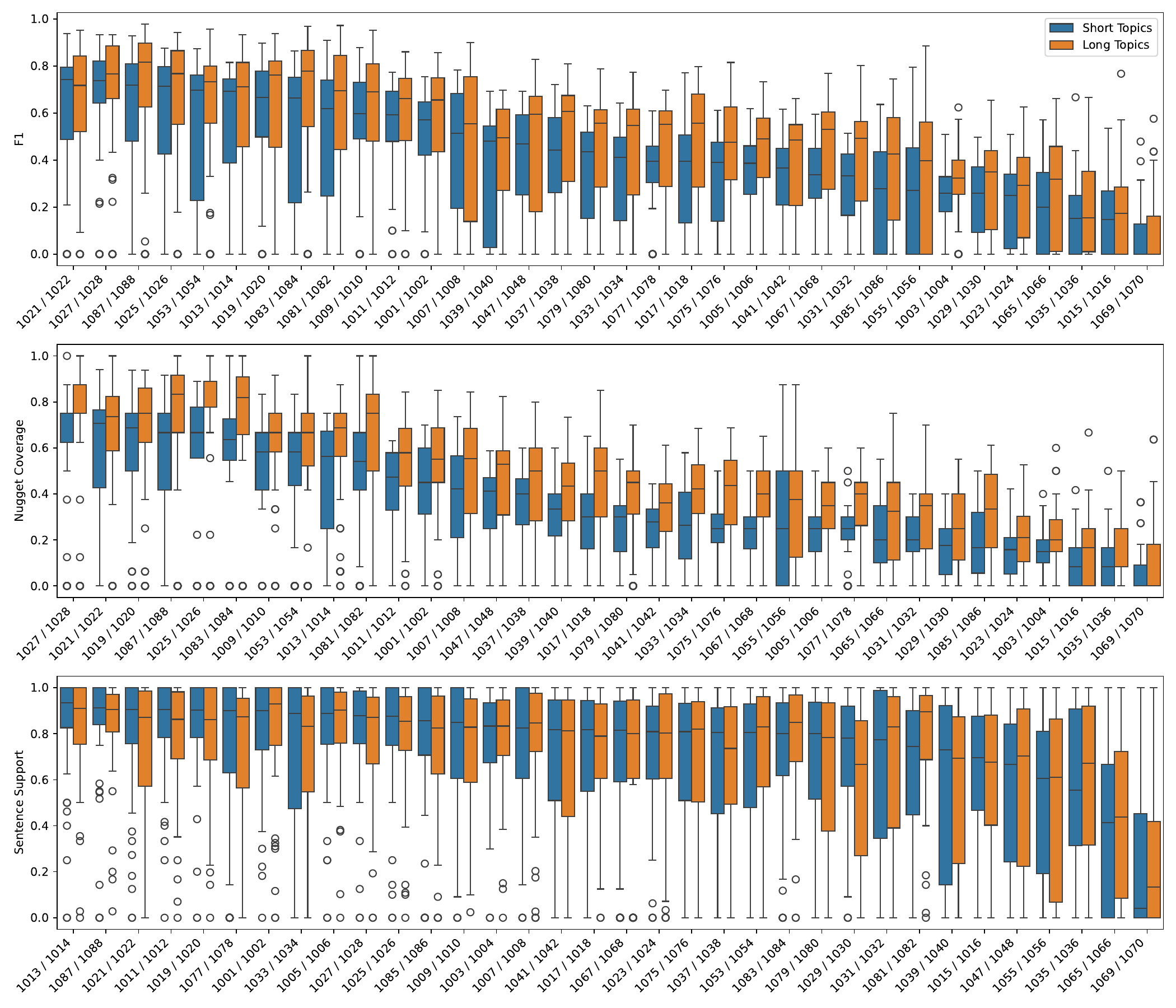}
    \caption{System Auto-ARGUE boxplots on pairs of short and long topics. Topic pairs are ordered by the median metric values of the short topics. }
    \label{fig:short-long-topics}
\end{figure*}

\eugene{
Another experiment in RAGTIME 2025 is the short and long topics. 
While it is possible to employ human assessors for reports under 2,000 characters, it would be cost-prohibitive to assess long reports. 
Figure~\ref{fig:short-long-topics} illustrates the score distribution of each topic on each of the three Auto-ARGUE metrics. 
While following the same difficulty trend, short and long topics exhibit different trends in score distributions. 
Intuitively, reports on long topics are able to cover more nuggets, likely because of the additional space. The faithfulness of reports is similar between short and long topics, with a larger portion of runs with lower sentence support scores across all topics, meaning that the longer the reports are, the more likely the systems start to hallucinate information. 
}

\eugene{
Table~\ref{tab:short-long-corr} further evaluates the similarity of the short and long topic evaluation results measured in correlation. 
While the Pearson correlation of the metric values on the topic-level seems to be high, indicating that the systems tend to translate the performance from short to long reports, the system ordering is less stable. 
However, it is worth noting that the system ordering correlation on the run level is similar to the correlation of human and Auto-ARGUE on the short topics, suggesting that the difference between short and long topics using Auto-ARGUE could be due to noise in the automatic evaluation implementation. 
}

\subsection{Multilingual Retrieval}

As an auxiliary task, RAGTIME still hosted a retrieval evaluation both for pool enrichment as well as our continued interest in evaluating multilingual retrieval.
Table~\ref{tab:mlir-results} summarizes the evaluation results of the 46 submissions. 
Since the document pools are shared across retrieval and report generation tasks, the cutoff is shallower than the ones we selected in NeuCLIR. 
However, all runs (with high priorities) have at least 16 documents judged in the top 20 (Judged@20 > 0.847). 

The runs consist of a wide range of retrieval methods, including BM25, learned sparse, dense, and LLM rerankers. 
These systems are a good representation of the current state-of-the-art in adhoc retrieval models. 
The top runs \texttt{hltime-qwen-jina} is a pipeline system that fuses three first-stage retrieval results (PLAID-X, LSR, and Qwen3 embeddings), followed by Qwen3 and Jina rerankers. 
The gap between the second and the third place is notably large (0.561 and 0.526), indicating Qwen3 reranker being substantially more effective than others. 

\section{Future Directions}
In 2026, we will continue with the RAGTIME track
with the primary task of report generation from multilingual news content in Spanish, Chinese, English, and Russian, removing the monolingual task.
In addition to the two remaining tasks, RAGTIME 2026 will add a new supporting task on autonuggetization that assesses how well participating systems can generate questions that should be answered in a report.

\subsection{What's New in RAGTIME 2026?}
RAGTIME 2026 will add a new Autonuggetization task, which will assess how well a system is able to identify the main topics that should appear in a report.
This capability underlies both the retrieval needed for report generation
and the structuring of the report.
The Autonuggetization task will directly measure this ability, whereas currently it is only measured implicitly in evaluations of report generation.
Given the retrieval collection and a report request,
participating systems will generate a list of $k$ single-sentence questions that should be answered in a report.
Assessors will align these questions with nuggets from the report generation task. 
Submissions will be evaluated based on how many nuggets they are aligned with.

\subsection{What's Not Changing?}
The Multilingual Report Generation and Multilingual Information Retrieval tasks will continue in RAGTIME 2026.
These tasks, and the new Autonuggetization task, will continue using a modified RAGTIME document collection where Spanish documents are substituted for Arabic documents.

\section{Conclusion}

RAG is now multilingual at TREC with RAGTIME! In this first year of RAGTIME we have worked together to forge a research community and
created new reusable evaluation resources.
Thirteen participating teams  contributed a total of 125 runs,
showing the citation accuracy is quickly becoming a solved problems as long as systems include reasonable checks,
but nugget coverage continues to be a work in progress.
Moreover, it is becoming apparent that retrieval has an important role to play in RAG
and that more work in retrieving diverse information is warranted.
The RAGTIME track will continue at TREC 2026,
perhaps with one or more additional tasks, so we have much to look forward to.

\bibliographystyle{ACM-Reference-Format}
\bibliography{biblio}

\appendix

\begin{table*}[t]
    \centering
    \caption{Report Generation Multilingual Subtask Results on Short Topics.}
    \label{tab:repgen-multi-results}
    
\begin{tabular}{ll|ccc|ccc}
\toprule
 &  & \multicolumn{3}{c|}{AutoARGUE} & \multicolumn{3}{c}{Human$^*$} \\
        &        & Nugget   & Sentence & \multirow{2}{*}{F1} & Nugget   & Sentence & F1 \\
Team ID & Run ID & Coverage & Support  &                     & Coverage & Support  & (LLM-Filled) \\
\midrule
coordinators & extractive\_rag & 0.498 & 0.771 & 0.562 & 0.227 & 0.550 & \bf{0.145} \\
GenAIus & genaius-cluster & 0.393 & 0.764 & 0.493 & 0.218 & 0.521 & 0.137 \\
HLTCOE & cru-ablR-PlaidX- & 0.347 & 0.934 & 0.481 & 0.178 & \bf{0.643} & 0.127 \\
hltcoe-multiagt & gptr\_nt\_q3d3\_mt & 0.386 & 0.865 & 0.498 & 0.196 & 0.524 & 0.127 \\
hltcoe-multiagt & lg\_nt\_4q12r3l\_mt\_c & 0.461 & 0.902 & \bf{0.577} & 0.201 & 0.487 & 0.124 \\
hltcoe-multiagt & auto\_swarm\_mt & 0.409 & 0.929 & 0.526 & 0.190 & 0.511 & 0.123 \\
hltcoe-multiagt & lg\_nt\_4q12r3l\_natv\_c & 0.431 & 0.894 & 0.544 & 0.204 & 0.476 & 0.122 \\
HLTCOE & cru-ansR-PlaidX- & 0.353 & 0.920 & 0.482 & 0.169 & 0.572 & 0.118 \\
HLTCOE & cru-ablR-LSR- & 0.363 & 0.946 & 0.492 & 0.161 & 0.596 & 0.118 \\
HLTCOE & cru-ablR- & 0.362 & \bf{0.961} & 0.498 & 0.174 & 0.559 & 0.117 \\
ncsu-las & las\_ag\_round\_robin & 0.426 & 0.732 & 0.510 & \bf{0.232} & 0.312 & 0.116 \\
hltcoe-rerank & hltime-lg.searcher & 0.451 & 0.739 & 0.527 & 0.192 & 0.385 & 0.115 \\
hltcoe-multiagt & gptr\_nt\_q4d4\_mt & 0.423 & 0.826 & 0.527 & 0.184 & 0.509 & 0.114 \\
AMU & AMU1ML & 0.358 & 0.837 & 0.450 & 0.181 & 0.505 & 0.113 \\
AMU & AMU1ENG & 0.397 & 0.789 & 0.478 & 0.178 & 0.496 & 0.113 \\
hltcoe-rerank & hltime-lg.crux.jsonl & 0.456 & 0.776 & 0.542 & 0.183 & 0.441 & 0.112 \\
hltcoe-multiagt & lg\_e2\_3q5r3l & 0.412 & 0.832 & 0.522 & 0.174 & 0.456 & 0.111 \\
hltcoe-rerank & hltime-lg.fsrrfprf & 0.452 & 0.768 & 0.540 & 0.169 & 0.422 & 0.110 \\
hltcoe-rerank & hltime-lg.jina & 0.488 & 0.756 & 0.548 & 0.181 & 0.375 & 0.107 \\
HLTCOE & cru-ansR- & 0.339 & 0.940 & 0.466 & 0.157 & 0.549 & 0.106 \\
HLTCOE & cru-ablR-conf- & 0.330 & 0.943 & 0.456 & 0.153 & 0.547 & 0.105 \\
HLTCOE & cru-ansR-conf- & 0.351 & 0.946 & 0.475 & 0.155 & 0.541 & 0.105 \\
hltcoe-rerank & hltime-lg.fsrrf & 0.444 & 0.756 & 0.527 & 0.164 & 0.406 & 0.104 \\
HLTCOE & cru-ansR-LSR- & 0.345 & 0.935 & 0.477 & 0.152 & 0.524 & 0.102 \\
hltcoe-rerank & hltime-gpt5.searcher & \bf{0.500} & 0.660 & 0.508 & 0.176 & 0.338 & 0.102 \\
hltcoe-rerank & hltime-lg.jina.qwen & 0.446 & 0.724 & 0.517 & 0.173 & 0.339 & 0.099 \\
HLTCOE & cru-ansR-bareconf- & 0.350 & 0.827 & 0.455 & 0.144 & 0.487 & 0.097 \\
hltcoe-multiagt & gptr\_e2\_q3d3\_mt & 0.356 & 0.853 & 0.473 & 0.141 & 0.499 & 0.096 \\
hltcoe-rerank & hltime-lg.qwen & 0.461 & 0.738 & 0.532 & 0.152 & 0.361 & 0.094 \\
ncsu-las & las\_ag\_sel\_28 & 0.447 & 0.597 & 0.470 & 0.173 & 0.278 & 0.090 \\
ncsu-las & las\_ag\_sel\_29 & 0.466 & 0.616 & 0.492 & 0.170 & 0.277 & 0.089 \\
hltcoe-rerank & hltime-lg.listllama & 0.448 & 0.709 & 0.508 & 0.154 & 0.368 & 0.089 \\
hltcoe-multiagt & gptr\_ka\_q3d3\_mt & 0.362 & 0.796 & 0.459 & 0.138 & 0.394 & 0.085 \\
IDACCS & IDACCS\_extract\_4.1\_combined & 0.365 & 0.646 & 0.413 & 0.127 & 0.469 & 0.084 \\
IDACCS & IDACCS\_hybrid\_4.1\_combined & 0.320 & 0.732 & 0.388 & 0.122 & 0.439 & 0.084 \\
hltcoe-multiagt & gptr\_ka\_q3d3\_natv & 0.369 & 0.763 & 0.471 & 0.122 & 0.417 & 0.083 \\
ncsu-las & ag\_sel\_new\_prompt & 0.458 & 0.660 & 0.517 & 0.129 & 0.274 & 0.076 \\
IDACCS & IDACCS\_hybridtb\_4.1\_combined & 0.300 & 0.774 & 0.372 & 0.108 & 0.432 & 0.074 \\
ncsu-las & ag\_sel\_all\_4.1 & 0.429 & 0.608 & 0.466 & 0.128 & 0.239 & 0.071 \\
GenAIus & genaius-question & 0.478 & 0.653 & 0.513 & 0.124 & 0.265 & 0.071 \\
IDACCS & IDACCS\_nugget\_4.1\_combined & 0.410 & 0.577 & 0.410 & 0.118 & 0.234 & 0.063 \\
IDACCS & IDACCS\_nuggetstb\_4.1\_combined & 0.365 & 0.553 & 0.372 & 0.104 & 0.223 & 0.061 \\
hltcoe-multiagt & lg\_e2\_3q5r2l\_mt\_qw3 & 0.282 & 0.334 & 0.241 & 0.069 & 0.145 & 0.038 \\
CSU & v3\_surround\_glm4 & 0.237 & 0.183 & 0.183 & 0.036 & 0.127 & 0.025 \\
CSU & v2\_split\_qwen & 0.256 & 0.156 & 0.116 & 0.025 & 0.102 & 0.015 \\
HLTCOE & cru-ansR-mostcommon- & --- & --- & --- & 0.012 & 0.206 & 0.006 \\
CSU & v1\_qwen & 0.223 & 0.004 & 0.005 & 0.005 & 0.038 & 0.004 \\
\bottomrule
\end{tabular}

\end{table*}

\begin{table*}[t]
    \centering
    \caption{Report Generation English Subtask Results on Short Topics.}
    \label{tab:repgen-eng-results}

\begin{tabular}{ll|ccc|ccc}
\toprule
 &  & \multicolumn{3}{c|}{AutoARGUE} & \multicolumn{3}{c}{Human$^*$==} \\
        &        & Nugget   & Sentence & \multirow{2}{*}{F1} & Nugget   & Sentence & F1 \\
Team ID & Run ID & Coverage & Support  &                     & Coverage & Support  & (LLM-Filled) \\
\midrule

WueRAG & WueRAG\_2025\_08\_22 & 0.365 & 0.797 & 0.466 & 0.152 & 0.530 & 0.107 \\
DUTH\_XANTHI & duth-mlir-tblocal-report & 0.064 & 0.505 & 0.103 & 0.024 & 0.383 & 0.019 \\
DUTH\_XANTHI & duth-mlir-mlm6-report & 0.066 & 0.491 & 0.095 & 0.020 & 0.378 & 0.016 \\
DUTH\_XANTHI & duth-mlir-mlm6loc-report & 0.066 & 0.491 & 0.095 & 0.020 & 0.386 & 0.016 \\
DUTH\_XANTHI & duth-mlir-xenc-report & 0.066 & 0.491 & 0.095 & 0.016 & 0.386 & 0.012 \\
DUTH\_XANTHI & duth-mlir-fused-report & 0.009 & 0.505 & 0.017 & 0.003 & 0.383 & 0.003 \\
DUTH\_XANTHI & duth-mlir-elec-report & 0.002 & 0.475 & 0.003 & 0.000 & 0.388 & 0.000 \\
DUTH\_XANTHI & duth-mlir-mlm12-report & 0.017 & 0.488 & 0.024 & 0.000 & 0.387 & 0.000 \\
DUTH\_XANTHI & duth-mlir-pybm25-report & 0.009 & 0.971 & 0.017 & 0.000 & 0.475 & 0.000 \\
DUTH\_XANTHI & duth-mlir-rrf-report & 0.000 & 0.480 & 0.000 & 0.000 & 0.350 & 0.000 \\
DUTH\_XANTHI & duth-mlir-tb-report & 0.001 & 0.466 & 0.003 & 0.000 & 0.362 & 0.000 \\
UvA & zetaalpha-english-gpt4omini-v1 & 0.374 & 0.153 & 0.160 & --- & 0.000 & --- \\
\bottomrule
\end{tabular}

\end{table*}

\begin{table*}[t]
    \centering
    \caption{Report Generation Results on Long Topics using Auto-ARGUE. ``E'' and ``M'' in the subtask column indicates the English and the multilingual subtasks, respectively. }
    \label{tab:repgen-results-long}

\begin{tabular}{ll|c|ccc}
\toprule
        &        &         & Nugget   & Sentence & \multirow{2}{*}{F1} \\
Team ID & Run ID & Subtask & Coverage & Support  &                     \\
\midrule
hltcoe-rerank & hltime-gpt5.searcher & M & 0.638 & 0.805 & 0.694 \\
GenAIus & genaius-question & M & 0.644 & 0.758 & 0.678 \\
hltcoe-multiagt & lg\_nt\_4q12r3l\_mt\_c & M & 0.535 & 0.918 & 0.647 \\
hltcoe-multiagt & lg\_nt\_4q12r3l\_natv\_c & M & 0.536 & 0.905 & 0.639 \\
hltcoe-rerank & hltime-lg.crux.jsonl & M & 0.555 & 0.792 & 0.636 \\
hltcoe-rerank & hltime-lg.searcher & M & 0.597 & 0.759 & 0.634 \\
HLTCOE & cru-ansR-conf- & M & 0.514 & 0.929 & 0.634 \\
HLTCOE & cru-ablR-conf- & M & 0.502 & 0.943 & 0.629 \\
HLTCOE & cru-ansR- & M & 0.506 & 0.936 & 0.623 \\
hltcoe-multiagt & gptr\_nt\_q4d4\_mt & M & 0.533 & 0.848 & 0.622 \\
hltcoe-multiagt & gptr\_nt\_q3d3\_mt & M & 0.517 & 0.859 & 0.617 \\
HLTCOE & cru-ansR-LSR- & M & 0.495 & 0.924 & 0.614 \\
HLTCOE & cru-ablR- & M & 0.488 & 0.953 & 0.611 \\
HLTCOE & cru-ansR-PlaidX- & M & 0.488 & 0.919 & 0.607 \\
HLTCOE & cru-ablR-PlaidX- & M & 0.480 & 0.943 & 0.605 \\
hltcoe-rerank & hltime-lg.listllama & M & 0.548 & 0.768 & 0.602 \\
hltcoe-multiagt & auto\_swarm\_mt & M & 0.489 & 0.945 & 0.597 \\
HLTCOE & cru-ablR-LSR- & M & 0.472 & 0.936 & 0.596 \\
hltcoe-rerank & hltime-lg.jina.qwen & M & 0.526 & 0.757 & 0.588 \\
IDACCS & IDACCS\_hybrid\_4.1\_combined & M & 0.506 & 0.792 & 0.587 \\
ncsu-las & las\_ag\_round\_robin & M & 0.573 & 0.651 & 0.586 \\
GenAIus & genaius-cluster & M & 0.488 & 0.825 & 0.585 \\
hltcoe-rerank & hltime-lg.fsrrfprf & M & 0.507 & 0.801 & 0.585 \\
hltcoe-rerank & hltime-lg.fsrrf & M & 0.511 & 0.741 & 0.584 \\
HLTCOE & cru-ansR-bareconf- & M & 0.504 & 0.770 & 0.581 \\
IDACCS & IDACCS\_hybridtb\_4.1\_combined & M & 0.521 & 0.740 & 0.579 \\
hltcoe-rerank & hltime-lg.jina & M & 0.511 & 0.748 & 0.577 \\
IDACCS & IDACCS\_extract\_4.1\_combined & M & 0.604 & 0.563 & 0.564 \\
coordinators & extractive\_rag & M & 0.498 & 0.774 & 0.562 \\
hltcoe-rerank & hltime-lg.qwen & M & 0.523 & 0.714 & 0.555 \\
hltcoe-multiagt & gptr\_e2\_q3d3\_mt & M & 0.440 & 0.825 & 0.548 \\
hltcoe-multiagt & lg\_e2\_3q5r3l & M & 0.440 & 0.848 & 0.540 \\
ncsu-las & ag\_sel\_all\_4.1 & M & 0.602 & 0.534 & 0.537 \\
ncsu-las & ag\_sel\_new\_prompt & M & 0.554 & 0.585 & 0.535 \\
hltcoe-multiagt & gptr\_ka\_q3d3\_mt & M & 0.458 & 0.742 & 0.528 \\
ncsu-las & las\_ag\_sel\_29 & M & 0.556 & 0.550 & 0.527 \\
AMU & AMU1ENG & M & 0.447 & 0.789 & 0.525 \\
ncsu-las & las\_ag\_sel\_28 & M & 0.582 & 0.537 & 0.523 \\
IDACCS & IDACCS\_nugget\_4.1\_combined & M & 0.604 & 0.499 & 0.523 \\
hltcoe-multiagt & gptr\_ka\_q3d3\_natv & M & 0.464 & 0.645 & 0.506 \\
IDACCS & IDACCS\_nuggetstb\_4.1\_combined & M & 0.602 & 0.465 & 0.496 \\
WueRAG & WueRAG\_2025\_08\_22 & E & 0.384 & 0.844 & 0.493 \\
AMU & AMU1ML & M & 0.396 & 0.797 & 0.478 \\
hltcoe-multiagt & lg\_e2\_3q5r2l\_mt\_qw3 & M & 0.355 & 0.436 & 0.309 \\
CSU & v3\_surround\_glm4 & M & 0.431 & 0.230 & 0.278 \\
CSU & v2\_split\_qwen & M & 0.244 & 0.143 & 0.129 \\
UvA & zetaalpha-english-gpt4omini-v1 & E & 0.579 & 0.074 & 0.120 \\
DUTH\_XANTHI & duth-mlir-tblocal-report & E & 0.067 & 0.510 & 0.107 \\
DUTH\_XANTHI & duth-mlir-mlm6loc-report & E & 0.066 & 0.495 & 0.095 \\
DUTH\_XANTHI & duth-mlir-mlm6-report & E & 0.066 & 0.495 & 0.095 \\
DUTH\_XANTHI & duth-mlir-xenc-report & E & 0.066 & 0.495 & 0.095 \\
CSU & v1\_qwen & M & 0.218 & 0.033 & 0.034 \\
DUTH\_XANTHI & duth-mlir-mlm12-report & E & 0.017 & 0.495 & 0.024 \\
DUTH\_XANTHI & duth-mlir-pybm25-report & E & 0.009 & 0.971 & 0.017 \\
DUTH\_XANTHI & duth-mlir-fused-report & E & 0.009 & 0.505 & 0.017 \\
DUTH\_XANTHI & duth-mlir-elec-report & E & 0.002 & 0.475 & 0.003 \\
DUTH\_XANTHI & duth-mlir-tb-report & E & 0.001 & 0.475 & 0.003 \\
DUTH\_XANTHI & duth-mlir-rrf-report & E & 0.000 & 0.480 & 0.000 \\
\bottomrule
\end{tabular}

\end{table*}

\begin{table*}[t]
    \centering
    \caption{MLIR Subtask Results.}
    \label{tab:mlir-results}

\begin{tabular}{ll|c|cc|c}
\toprule
Team ID & Run ID & Judged@20 & MAP & R@1000 & nDCG@20 \\
\midrule
hltcoe-rerank & hltime-qwen-jina & 0.982 & 0.454 & 0.610 & 0.661 \\
hltcoe-rerank & hltime-qwen & 0.988 & 0.451 & 0.610 & 0.650 \\
hltcoe-rerank & hltime-listllama & 0.982 & 0.378 & 0.481 & 0.624 \\
hltcoe-rerank & hltime-searcher & 0.982 & 0.351 & 0.478 & 0.622 \\
hltcoe-rerank & hltime-fsrrfprf & 0.979 & 0.459 & 0.801 & 0.607 \\
hltcoe-rerank & hltime-fsrrf & 0.982 & 0.449 & 0.796 & 0.602 \\
GenAIus & genaius-llama3-3-70B & 0.974 & 0.359 & 0.640 & 0.589 \\
hltcoe-rerank & hltime-rankk & 0.979 & 0.285 & 0.396 & 0.587 \\
GenAIus & genaius-gpt-4o & 0.975 & 0.375 & 0.642 & 0.585 \\
GenAIus & genaius-gpt-oss-20b & 0.979 & 0.295 & 0.592 & 0.543 \\
hltcoe-rerank & hltime-fsqwen & 0.987 & 0.364 & 0.722 & 0.535 \\
hltcoe-rerank & hltime-lsr & 0.982 & 0.319 & 0.685 & 0.512 \\
GenAIus & genaius-gpt-oss-120b & 0.982 & 0.256 & 0.560 & 0.498 \\
hltcoe-rerank & hltime-plaidx & 0.982 & 0.305 & 0.537 & 0.496 \\
coordinators & bm25-td-rank1 & 0.957 & 0.218 & 0.551 & 0.417 \\
coordinators & bm25-t-rank1 & 0.943 & 0.205 & 0.490 & 0.406 \\
coordinators & bm25-d-rank1 & 0.963 & 0.192 & 0.507 & 0.399 \\
coordinators & mt-bm25-td & 1.000 & 0.134 & 0.232 & 0.357 \\
coordinators & mt-bm25-title & 1.000 & 0.115 & 0.210 & 0.297 \\
coordinators & bm25-d-rankk & 1.000 & 0.067 & 0.097 & 0.214 \\
DUTH\_XANTHI & duth\_mlir\_xenc & 0.375 & 0.041 & 0.095 & 0.164 \\
DUTH\_XANTHI & mlir-pybm25 & 0.375 & 0.041 & 0.095 & 0.164 \\
DUTH\_XANTHI & mlir-fused & 0.372 & 0.041 & 0.092 & 0.163 \\
DUTH\_XANTHI & mlir-tblocal & 0.418 & 0.030 & 0.092 & 0.114 \\
DUTH\_XANTHI & duth-mlir-mlm6 & 0.419 & 0.031 & 0.092 & 0.114 \\
DUTH\_XANTHI & duth-mlir-mlm6loc & 0.419 & 0.031 & 0.092 & 0.114 \\
DUTH\_XANTHI & mlir-tb & 0.309 & 0.024 & 0.092 & 0.086 \\
coordinators & bm25-t-rankk & 1.000 & 0.016 & 0.027 & 0.071 \\
coordinators & bm25-td-rankk & 1.000 & 0.016 & 0.027 & 0.071 \\
DUTH\_XANTHI & duth\_mlir\_eng\_rrf & 0.322 & 0.018 & 0.092 & 0.061 \\
DUTH\_XANTHI & mlir-mlm12 & 0.244 & 0.010 & 0.092 & 0.038 \\
DUTH\_XANTHI & mlir-elec & 0.218 & 0.009 & 0.092 & 0.031 \\
\bottomrule
\end{tabular}

\end{table*}

\end{document}